# Vector Embeddings by Sequence Similarity and Context for Improved Compression, Similarity Search, Clustering, Organization, and Manipulation of cDNA Libraries

Daniel H. Um[a,*], David A. Knowles[b,c], Gail E. Kaiser[d]

[a]*Department of Computer Science, Columbia University, New York, NY, USA*
[b]*Department of Computer Science, Columbia University, New York, NY, USA*
[c]*New York Genome Center, New York, NY, USA*
[d]*Department of Computer Science, Columbia University, New York, NY, USA*

[*]Corresponding author: Daniel Um, dhu2102@columbia.edu

*Abstract*—This paper demonstrates the utility of organized numerical representations of genes in research involving flat string gene formats (i.e., FASTA/FASTQ[5]). FASTA/FASTQ files have several current limitations, such as their large file sizes, slow processing speeds for mapping and alignment, and contextual dependencies. These challenges significantly hinder investigations and tasks that involve finding similar sequences. The solution lies in transforming sequences into an alternative representation that facilitates easier clustering into similar groups compared to the raw sequences themselves. By assigning a unique vector embedding to each short sequence, it is possible to more efficiently cluster and improve upon compression performance for the string representations of cDNA libraries. Furthermore, through learning alternative coordinate vector embeddings based on the contexts of codon triplets, we can demonstrate clustering based on amino acid properties. Finally, using this sequence embedding method to encode barcodes and cDNA sequences, we can improve the time complexity of the similarity search by coupling vector embeddings with an algorithm that determines the proximity of vectors in Euclidean space; this allows us to perform sequence similarity searches in a quicker and more modular fashion.

## 1. Introduction & Literature Review

As the cost of sequencing DNA falls at a rate comparable to Moore's Law, many parallels have been drawn between computer science and synthetic biology, a new multidisciplinary field.

If the four base pairs of DNA are likened to the cell's binary 1s and 0s, the ribosome would be the cell's compiler, and novel technologies such as CRISPR-Cas9[1] and next-generation sequencing would provide cellular read and write privileges. In recent years, much research in synthetic biology has focused on developing tools that researchers can use to program cells. The art of genetic data management[4], on the other hand, has been slow to develop. This is partly due to a lack of consensus on how genetic researchers should organize, store, and access data. Due to this lack of centralization for genetic research software tools[3,7,9,14,33,34], an average of one and a half months is required to even replicate the results of previous bioinformatic research [35], because of the way the data is organized, stored and accessed causing computation to be slow.

This investigation was primarily motivated by the corresponding author's experience working as a bioinformatics analyst for Fluent BioSciences[2] and in his academic computational biology research. Industry norms for working with genomic data files are clumsy and inefficient. For example, during work with single-cell RNA-seq data, runs need to be stored in AWS S3 Drive folders arranged by run date and read into Python files by path using a method that reads-in files line-by-line, with some stripping and extra checks.

---

[1] Commonly referred to as "genetic scissors", CRISPR-Cas9 is a gene-editing technology that enables precise modifications to DNA by cutting and replacing specific segments of genetic material.
[2] https://www.fluentbio.com/

Genes are rarely saved as objects; instead, typical pipelines are used to generate quantifications from flat files. In addition, several inefficiencies are to be noted in the process of data handling[29] that could be improved with the use of software database[1] methods and principles, specifically implementing compression algorithms to reduce storage requirements, enabling efficient similarity searches to identify related cDNA sequences, optimizing clustering techniques for categorizing genes with similar expression patterns, establishing effective data organization methods for easy retrieval and management of cDNA libraries, and developing versatile manipulation tools for seamless analysis and exploration of gene expression data within the libraries.

The Childhood Cancer Data Lab in Philadelphia[3] provided an ensemble method for the reduction of dimensionality (which can help reduce the computational expense of vector-embeddings in high dimensions),[40] Tang et al. described the benefit of similarity-based clustering before compression[12], and Minicom showed a novel compression method.[37] Du et al. presented a vector-embedding method for distributed gene co-expression in which they utilize transcriptome-wide[17] gene co-expression data to predict gene-gene interaction, based solely on gene names. A different means of vector embeddings in working with cDNA libraries without the need for gene classification is described in the Tang et al paper[12]. The findings of these four key literature review papers will be elaborated in the Methods section, particularly how they initiated this paper's investigation and direction.

## 2. METHODS

*2.1 Overview*

With respect to the datasets used in this investigation, context embeddings were trained on short-sequence Uniprot aggregate FASTA files with 3D structures, Swiss-Prot Reviewed Binding Site, and human alternative splicing while clustered compression[24] and similarity search[15] speed comparisons were performed on three FASTA files of approximately 20 MB in size constructed from experimental data and titled sandmouse.fa, moriarty-neg.fa, and pathogen.fa. Once the vector-embedding scripts were written and tested using these files, cDNA data from 10x Genomics were downloaded and used for the remainder of the investigation.[8]

The methodology for the DNA base-pair representation exploration was rooted in the idea that by clustering similar sequences together, determining similarity through each string's vector-embedding representation, it would be possible to improve the performance of tasks that depend on semantic ordering, such as similarity search[20], ranking/recommendation, deduplication/record matching, and anomaly detection. For data pre-processing, file readers were built to have the capacity of efficiently taking in multiple different FASTA types, including being able to deal with edge cases, such as like a mix of lowercase and uppercase strings (e.g., "AtCG") as well as FASTA files containing "none" reads and reads of vastly different lengths. The functions were written to encode and decode DNA strings into unique integers (e.g., using a 16^ index multiplied by a one-hot-coded base in bytes to encode and a modulus operator to peel layers back). Additionally, an altered form of k-means clustering was used for this implementation, in which similarity was based on two-mer frequency with the original indices preserved. Furthermore, clustered sequences were converted into byte arrays, where the delineators between sequences are the original indices, such that the structure enabled efficient conversion back into the original FASTA/FASTQ format if necessary.

An improved vector embedding based on context was written,[30] where triplets of nucleotides (i.e., amino acid codons) were abstracted as words in the context of the polypeptide sentence. The embedding was learned to accurately predict the center codon from its neighbors (as is show in the skip-gram model). In Figure 1, the center word is CUU at position t, and the window size is two, spanning from $w_{t-2}$ to $w_{t+2}$.

---

[3] https://www.chop.edu/cccr

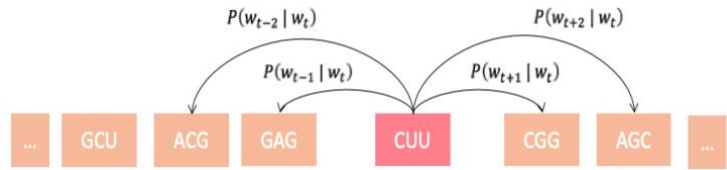

FIGURE 1: Skip-gram model showing center codon highlighted in red and a window size of two

Amino acids were mapped to one-hot-encoded vectors and fed in as inputs. Linear neurons, used as a lookup table, were the hidden layer, with weights in the hidden layer optimized using gradient descent, and new weights were back propagated for the hidden layer. The output layer was optimized using softmax to support cross-entropy from the output layer to the target layer (the current amino acid sequence from its context). The extracted weight matrices from the hidden layer were used to create the 2D context embeddings for each amino acid.

The sequence embedding, context embedding, barcode, original file index, codon information, gene information, and t-SNE data were centralized into a four-table relational database[21,25,39,41] for the easy reproduction of the results and access for future experimentation.[38] To demonstrate the ease of manipulation, a t-SNE plot color coded by vector-embedding values was produced—this plot is not usually produced but serves as a proof of concept to demonstrate the ease of joining two tables through a primary key (i.e., the barcode) and for rapid visualization.

Following the creation of vector-embedding algorithms and the organization of data, sequences were clustered in relation to the distances between sequence embeddings. The hypothesis for testing using this clustering method was that certain compression algorithms (ones that reduce size by consolidating redundancies[2,6,16,23,26,27,32,35]) can be improved by compressing clusters of similar sequences. Another application of the vector embeddings was the similarity search. Tools from FAISS, Facebook Research's library for efficient similarity search and clustering of dense vectors, were utilized to perform speed benchmarks. However, for these tests, each integer vector embedding needed to be distributed into a nx64 embedding float32 array, and cDNA barcodes needed to be mapped to a FAISS index. IndexFlatL2 was used as the distance metric.[22]

*2.2 Sequence Embedding*

The first type of vector embedding was constructed as follows. The characters A, T, C, and G were mapped to unique integer representations. Different mappings were tested, including binary-encoded integer representation, where the integer versions of 1000, 0100, 0010, and 0000 were used to represent A, T, C, and G; the standard mapping, where 0, 1, 2, and 3 were used to represent the nucleotides; and no encoding, where the binary representations of the characters themselves were used. Although the specific code was written in Python, for generalizability, pseudocode is presented here.

```
SET bits from [2,4,8];
SET base_size TO 2^bits;
if encoding EQUALS onehotencoding then
    SET onehotencoding._mapping TO
        "A": 8, "C": 4, "G": 2, "T": 1, "N": 15, "E": 0;
    SET onehotencoding._rmapping TO
        8: "A", 4: "C", 2: "G", 1: "T", 15: "N", 0: "E";
end
else if encoding EQUALS standard then
    SET standard._mapping TO
        "A": 0, "C": 1, "G": 2, "T": 3, "N": 4;
    SET standard._rmapping TO
        0: "A", 1: "C", 2: "G", 3: "T", 4: "N";
end
else
    SET noencoding._mapping TO
        No change from FASTA and FASTQ char dtypes
end
```
ALGORITHM 1: MAPPINGS FOR VECTOR EMBEDDINGS

After the characters were mapped to their integer representations, strings were encoded into their unique numerical forms, using Algorithm 2. This algorithm multiplies the integer mapping of each nucleotide base by an encoder (as defined in the previous mapping function; three bits were used for 2, 4, and 8 for the encoder).

Then, the numerical representation of each string was returned, such that each string had a unique representation, and the closer that two of these numerical representations were, the more similar the strings were in terms of their raw nucleotide sequence.

```
Data: sequence: string
Result: integer embedding
SET val TO 0;
for i, base IN enumerate(sequence) do
    val += encoding._mapping[base] * (base_size^i);
end
return val
```
ALGORITHM 2: ALGORITHM FOR ENCODE SEQUENCE AS UNIQUE INTEGER EMBEDDING

After each string was embedded into its unique numerical representations, it was converted into a byte array to be usable for compression and search. Algorithm 3 divided the unique integer representing each string and converted its smaller chunks into bytes.

```
Data: self, cluster: list of strings
Result: bytearray
SET ec TO [];
SET N TO len(cluster);
SET indices TO [];
SET lengths TO [];
SET byte_array TO N.to_bytes(4, "little");
for sequence, index IN cluster) do
    SET length TO
     ⌊(len(sequence) + (⌊8/self._bits⌋ − 1))/(8/self._bits)⌋;
    SET es TO self.encode_sequence(sequence).to_bytes(length, "little");
    ec.append(es);
    indices.append(index);
    lengths.append(length);
end
for index IN indices) do
    byte_array += index.to_bytes(4, "little");
end
for length IN lengths) do
    byte_array += index.to_bytes(1, "little");
end
for es in ec) do
    byte_array += es;
end
return byte_array
```

ALGORITHM 3: ALGORITHM FOR ENCODING CLUSTER AS VECTOR EMBEDDINGS

Note that this step involves a slight increase in size; however, the performance gain from compressing clustered byte arrays based on sequence similarity outweighed the temporary increase.

Furthermore, to reconstruct the original FASTA or FASTQ file from this compressed byte array, we needed to store the indices as delineators between subarrays. This was the primary reason for the slight increase in size between the numerical integer representations for the strings and the byte array. However, because we mixed the order of the sequences in the FASTA or FASTQ file to cluster and maximize compression, this was a necessary step. Furthermore, although it was not relevant for a cDNA library where barcodes are used, with plain FASTA files, the order is important for concatenating short reads into larger genes.

Ultimately, the construction of these vector embeddings for gene sequences in the form of byte arrays were important for compression and similarity search, as is discussed further later in this paper.

*2.3 Clustering*

Before moving on to compression, it is important to explain the theory behind the proposed method of improving the compression ratio. First, it is important to note previous attempts to improve the industry standard of genetic data compression.

Mansouri et al.[28] published a novel binary method of encoding genes, in which genetic information is stored in four files, encoding using a recursive decision step that encodes the base pair (A, T, C, or G) that has the highest frequency as one and the rest as zeros, stores this as one file, and compresses the file. The DNA-SBE system "outperforms state-of-the-art compressors and proves its efficiency in terms of special conditions imposed on compression data," according to the paper, which claims that it also "outperforms state-of-the-art compressors and proves its efficiency in terms of special conditions imposed on compression data."[28] The Mansouri et al. article contains extremely useful graphics and pseudocode that allow the reader to comprehend the reasoning and recreate the study.

The Broad Institute study, as referenced in the introduction, identified several useful unsupervised algorithms that could improve biological representations that utilize characteristics collected from various compression models. The ensemble method outlined compressed more effectively and provided a more nuanced approach to dimensionality reduction NMF (non-negative matrix factorization) than traditional techniques such as principal component analysis) and NMF. Furthermore, Tang et al. proposed proposes that in unsupervised compression of short read databases, using clustering based on vector embeddings that are rooted in sequence similarity, one can losslessly improve certain compression methods.

Our method of clustering is a slightly altered version of vector-embedding k-means clustering that clusters granularly into groups of three, before clustering larger chunks. This allows modularity in the clusters that are compressed, which is helpful because certain compression algorithms perform better with larger clusters, and others work better with tight clusters of highly similar sequences.[30] The commonality between compression methods that work best using this method of clustering is reducing file size by addressing redundancies.

In addition to the use of the algorithm given below, our method involves selecting the features that have the highest variance to inform groups of clusters to compress together.

```
Data: V: array, l: Optional[float] = 0.3
Result: array
if len(V) < 3 then
    RETURN [V]
end
else
    SET c1, c2 TO kmeans(V, 2) if len(c1) EQUALS 0 then
        RETURN [c2]
    end
    if len(c2) EQUALS 0 then
        RETURN [c1]
    end
    if len(c1) + len(c2) ≥ 4 and (len(c1) EQUALS 1 and len(c2)
      EQUALS 1) then
        if len(c1) ≠ 1 then
            SET c2, c1 TO c1, c2 SET ca, cb TO c1, c2
        end
        SET cbmin TO mindist(ca[0], cb)
        if dist(ca[0], cbmin) < l * dist(mindist(cbmin, cb[cb ≠ cbmin]))
          then
            SET ca TO np.array([ca[0], cbmin])
            SET cb TO cb[cb ≠ cbmin]
            RETURN cluster(cb, l) + [ca]
        end
    end
    else
        RETURN cluster(c1, l) + cluster(c2, l)
    end
end
```

ALGORITHM 4: CLUSTERING

## 3. RESULTS

*3.1 Context Learning*

To identify some meaning in the vector embeddings of genes, we can take inspiration from natural language processing. In these vector embeddings, we can treat codons as words and sequences as sentences. To convert raw sequences into codons, nucleotide triplets were assigned to their one-letter amino acid representations or to the letter "O" if their triplet does not map to any one amino acid. In all, three frames were tested per short sequence, and the frame that had the lowest number of non-assignments was chosen.

$$L(\theta) = \prod_{k=1}^{T} \prod_{-m<J<m} Pr(w_{k+j}|w_k, \theta)$$

Further, we define the loss function as the minimization of a negative likelihood. In the following equation, m represents the window size (in the given case, 2).

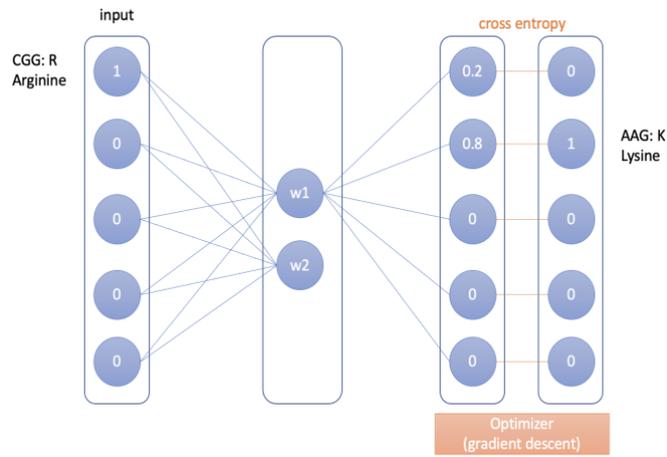

FIGURE 2: Model for learning weights

As can be seen in Figure 2, he one-hot-encoded version of the codon was initially fed in as the input, and then in the hidden layer, the weights are given initial determination, followed by gradient descent and cross-entropy with the target to update the weights in the hidden layer, as can be seen in the diagram above.

Although this did not occur very often, particularly regarding training on a smaller subset of the data, for the use of a larger training set, the model likely encounters more non-mapping amino acids that do not follow a pattern (because they are simply a catch-all). In our scripts, we omit non-mapping amino acids from the overall table and the graph; however, if non-mapping amino acids are plotted, this moves the result further from the remaining amino acids as more data is trained, which pushes the remaining amino acids closer together. One way around this would be to exclude any short read that includes a non-mapping amino acid; however, this would inherently exclude any short read that has a single nucleotide misread, and we would therefore be discarding important data. Another way of addressing this issue would be not to include the data in the model itself, but this would exclude training on non-mapped amino acids and the amino acid(s) that come after these non-mapped amino acids, in a way that is dependent on the window size used.

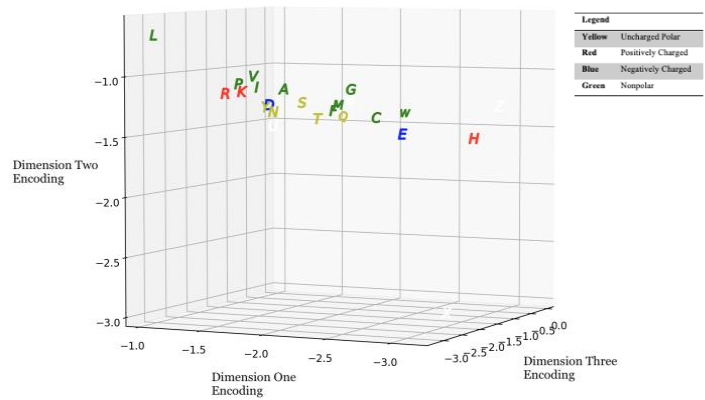

FIGURE 3: Context embeddings trained on short-sequence Uniprot FASTA files with **3D structures** – N = 17552

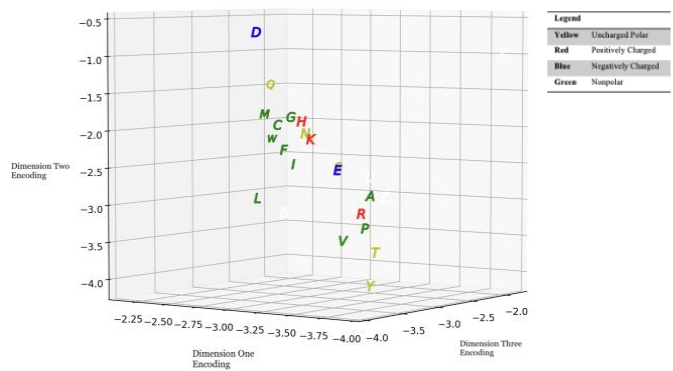

FIGURE 4: Context embeddings trained on short-sequence Uniprot FASTA files for **Swiss-prot reviewed binding site** – N = 28,052

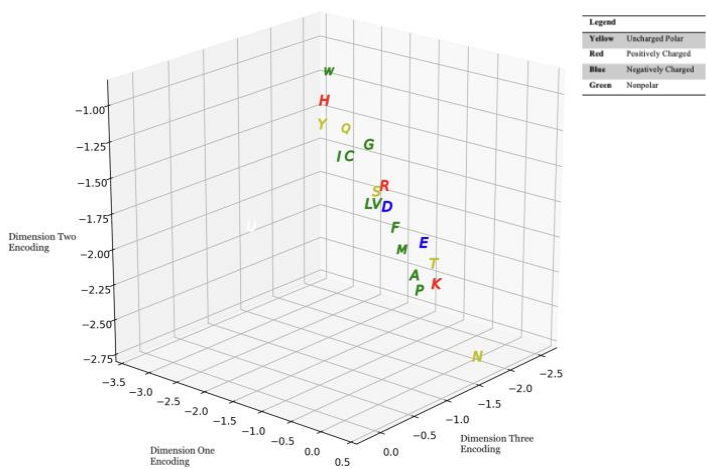

FIGURE 5: Context embeddings trained on short-sequence Uniprot FASTA files for **human alternative splicing** – n = 10,541

As noted in the Methods section, amino acids were mapped to one-hot-encoded vectors and fed in as the inputs. A linear neuron, used as a lookup table, was used as the hidden layer, weights in the hidden layer were optimized using gradient descent, and new weights were back propagated in the hidden layer. The output layer was optimized using softmax.

As can be seen in Figures 3-5, patterns of clustering appear to emerge between amino acids with similar properties. Furthermore, the dataset of the 3D structures and the binding site dataset seem to show similar embeddings (i.e., proline, valine, alanine, arginine, and tyrosine cluster together, as do methionine, cysteine, glycine, tryptophan, phenylaniline, and glutamine, etc.) However, the alternative splicing dataset seems to learn substantially different embeddings, which is likely because, by definition, sequences in this dataset are entropically not the most optimal.

## 3.2 Clustered Compression

The initial compression experiments were performed on three FASTA files, pathogen.fa, moriarty-neg.fa, and sandmous.fa, each of which was approximately 20 MB. The compression algorithm used for the preliminary tests was gzip, as it is built into Python and is the most common compression algorithm currently used for FASTA files.[31]

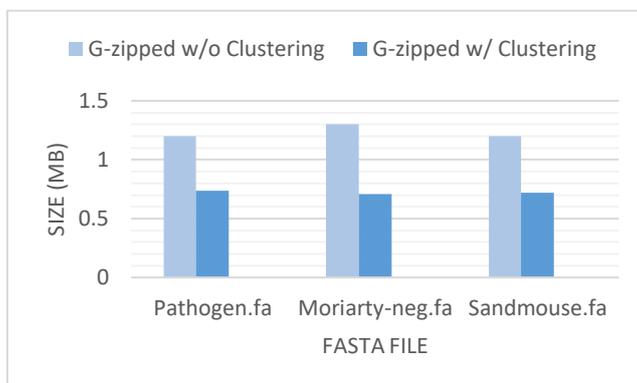

FIGURE 6: Bar chart showing compression improvements on the FASTA files pathogen.fa, moriarty-neg.fa, and sandmous.fa, comparing gzip w/o clustering and size after gzip with clustering. Note, the original one-hot encoded FASTA file sizes were all ~24.3 MB for all three files.

As can be seen in Figure 6 above, compressing with clustering improved the compression performance by about 50%. It is interesting to note is that although moriarty-neg.fa showed the best clustered compression, it had the worst non-clustered performance. This results from the fact that the improvement in performance is dependent on the similarity of sequences in each cluster and the clusterability of the original dataset.

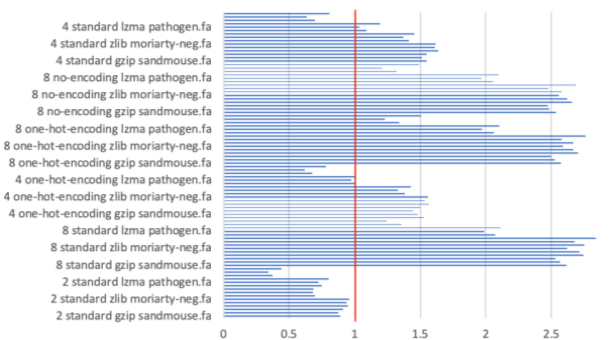

FIGURE 7: Compression ratio performance vs. bits per base, encoding, compressor on the three FASTA files pathogen.fa, moriarty-neg.fa, and sandmous.fa

A more extensive experiment was conducted, as shown in Figure 7, to examine a variety of bits per base (i.e., 2, 4, and 6), three encoding methods (i.e., one-hot-encoding, standard encoding, and no encoding), five compression algorithms (i.e., lzma, zlib and gzip, bz2, and zpac), with all permutations run on the three tester files, pathogen.fa, moriarty-neg.fa, and sandmous.fa. For more information on the aforementioned parameters, such as bits per base, encoding, and compressors, please refer to the Methods section.

The best performance was seen with 8 bits per base, in standard encoding, using lzma on pathogen.fa. Among the different parameters, the bits per base seemed to have the largest effect, and in particular, 4 bits per base seemed to have the largest variance, based on the coupled compressor. It is also noteworthy that one-hot-encoding performed almost as well as standard encoding, as we needed to add extra index information, where both were clustered by sequence similarity.

Additionally, although five compressors were tested, the results of only three are displayed in the graph (i.e., gzip, zlib, and lzma) for readability. With regard to the results, lzma performed the best and zpac performed by far the worst. However, when comparing gzip and lzma, the trade-off between speed and compression ratio is important to note, where, despite performing more quickly than gzip, the current industry standard, lzma does take around 10% longer to compress. For this reason, a researcher who requires rapid compression and decompression may decide to stick to gzip, while someone who is working with immense amounts of data that they would typically archive when not in use may find that compressing with lzma is the better solution. For the full table, with 90 rows of the complete compression experiment, please refer to the appendix.

*3.3 Similarity Search and Speed Comparisons*

Finally, vector embeddings were used to compare the speed of running similarity searches between DNA sequences. Before diving into our implementation of similarity searches using vector embeddings, it may be useful to explain what similarity search using vectors would entail on an abstract level.

First, this method treats our set of vectors $x_i$ in dimension d as an index, where computing the $argmin_i$ entails searching for the closest i indices, including itself. The advantages of vector-embedding similarity search over string similarity search, which is the current standard, is that this method can search several sequences simultaneously, rather than being restricted to a single one at a given time (i.e., batch processing). Vector-embedding similarity search can specify the specific ratio between precision and speed, and a vector-embedding similarity search can perform the maximum inner product search (i.e., $argmax_i(x, x_i)$) instead of the minimum Euclidean search. Furthermore, with the use of Facebook research's FAISS package, it is also possible to return all elements that are within a given radius of the query point (i.e., a range search) and store the index on a disk rather as RAM. The following equation provides vector-embedding similarity search using argmin Euclidean distance ($L^2$) with FAISS.

$$j = argmin_i ||x - x_i||$$

Jellyfish is a library that can be used for performing string similarity searches. It contains multiple comparison options, including the Levenshtein distance, the Damerau-Levenshtein distance, the Jaro distance, the Jaro-Winkler distance, match rating approach comparison, and Hamming distance. Practically speaking, these different metrics can work interchangeably, and the primary difference between their potential use case is the way that each handles deletions, insertions, and substitutions with respect to how each affects the overall similarity score. The following equation provides Jaro similarity with Jellyfish.

$$sim_j = \begin{cases} 0 \; if \; m = 0 \\ \frac{1}{3} \left( \frac{matches}{str_1} + \frac{matches}{str_2} + \frac{matches - transpositions}{matches} \right) otherwise \end{cases}$$

For the purposes of this experiment, the Jaro distance was used because it is rooted in a measure of characters in common, while the Levenshtein distance counts the number of edits needed to convert one string into the other, and the Hamming distance is the number of places where the characters are different. Therefore, there is a more direct comparison between the vector-embedding string similarity search and the Jaro distance than Levenshtein or Hamming, and the Jaro distance was used instead of Jaro-Winkler for simplicity.

| Sequences | FAISS (s) | Jellyfish(s) |
|---|---|---|
| 10 | 0.2105 | 0.0042 |
| 100 | 0.0762 | 0.0278 |
| 1000 | 0.0968 | 0.1308 |
| 10000 | 0.0181 | 1.2905 |
| 100000 | 0.0213 | 12.8318 |

TABLE 1: TABLE FOR SPEED COMPARISONS BETWEEN FAISS AND JELLYFISH (STRING SIMILARITY SCORING PACKAGE)

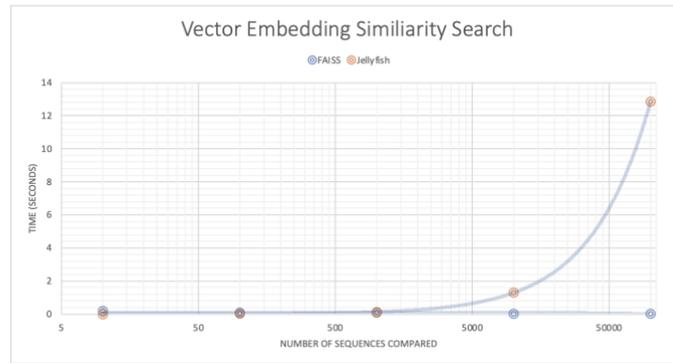

FIGURE 8: Speed comparison of FAISS with Jellyfish (string similarity scoring package)

As can be seen in the results of this experiment, in Table 1 and Figure 8, although the Jellyfish string similarity search is initially performed between smaller datasets, it does not scale well (exponentially) with number of sequences, relative to FAISS. By contrast, the vector-embedding similarity search, remains relatively constant with respect to time complexity.

The primary reason for the differential time complexities has to do with the batch processing of vector indices, by contrast to pairwise similarity search, as is the case with Jellyfish.

## 4. CONCLUSIONS

The value of the ordered numerical representations of genes in research employing flat string formats of genes is demonstrated in this work (i.e., FASTA, FASTQ, and cDNA libraries). We can cluster more efficiently and increase compression performance by assigning each short sequence a unique vector embedding. Furthermore, by encoding barcodes and cDNA sequences with this sequence embedding approach, we can create a relational database that, when combined with Facebook research's FAISS algorithm, allows us to perform sequence similarity searches more rapidly and with greater modularity than using industry standard methods. We may also demonstrate grouping based on amino acid characteristics by learning alternative coordinate vector embeddings, depending on contexts of codon triplets. Some possible extensions to be explored could be optimizing uncompressing to the original sequences, speed of mapping for similarity search, and downstream applications of amino acid learned embeddings.

## DECLARATION OF INTEREST

PROF. KAISER'S WORK IS FUNDED IN PART BY DARPA/NIWC-PACIFIC N66001-21-C-4018 AND IN PART BY NSF CCF-1815494, CNS–2247370 AND CCF-231305.

DANIEL UM DECLARES THAT HE HAS NO KNOWN COMPETING FINANCIAL INTERESTS OR PERSONAL RELATIONSHIPS THAT COULD HAVE APPEARED TO INFLUENCE HIS CONTRIBUTION TO THIS PAPER.

## DECLARATION OF GENERATIVE AI IN SCIENTIFIC WRITING

THIS RESEARCH DID NOT MAKE USE OF ANY GENERATIVE AI IN ITS WRITING OR PRODUCTION.

## ETHICS APPROVAL STATEMENT

THIS RESEARCH DID NOT CONTAIN ANY STUDIES INVOLVING ANIMAL OR HUMAN PARTICIPANTS, NOR DID IT TAKE PLACE ON ANY PRIVATE OR PROTECTED AREAS. NO SPECIFIC PERMISSIONS WERE REQUIRED FOR CORRESPONDING LOCATIONS.

## ACKNOWLEDGEMENTS

I would like to recognize Shouvik Mani, Jincheng Xu, and Leslie Ramos for proofreading the manuscript and for their friendship.

## MENDELEY DATA

Context embeddings trained on short sequence FASTA files with binding site on Mendeley Data are associated with "FIGURE 4: Context embeddings trained on short-sequence FASTA files with binding site – N = 28,052" for "Vector Embeddings by Sequence Similarity and Context for Improved Compression, Similarity Search, Clustering, Organization, and Manipulation of cDNA Libraries" by Daniel H. Um et al.

Context embeddings trained on short sequence FASTA files for human alternative splicing on Mendeley Data are associated with "FIGURE 5: Context embeddings trained on short-sequence FASTA files for human alternative splicing – n = 10,541" for "Vector Embeddings by Sequence Similarity and Context for Improved Compression, Similarity Search, Clustering, Organization, and Manipulation of cDNA Libraries" by Daniel H. Um et al.

Context embeddings trained on short sequence FASTA files with 3D structures on Mendeley Data are associated with "FIGURE 3: Context embeddings trained on short-sequence FASTA files with 3D structures – N = 17552" for "Vector Embeddings by Sequence Similarity and Context for Improved Compression, Similarity Search, Clustering, Organization, and Manipulation of cDNA Libraries" by Daniel H. Um et al.